

\frenchspacing

\parindent15pt

\abovedisplayskip4pt plus2pt
\belowdisplayskip4pt plus2pt
\abovedisplayshortskip2pt plus2pt
\belowdisplayshortskip2pt plus2pt

\font\twbf=cmbx10 at12pt
 at12pt
 at12pt

\font\sc=cmcsc10

\font\ninerm=cmr9
\font\nineit=cmti9
\font\ninesy=cmsy9
\font\ninei=cmmi9
\font\ninebf=cmbx9

\font\sevenrm=cmr7

\font\seveni=cmmi7
\font\sevensy=cmsy7

\font\fivenrm=cmr5
\font\fiveni=cmmi5
\font\fivensy=cmsy5

\def\nine{%
\textfont0=\ninerm \scriptfont0=\sevenrm \scriptscriptfont0=\fivenrm
\textfont1=\ninei \scriptfont1=\seveni \scriptscriptfont1=\fiveni
\textfont2=\ninesy \scriptfont2=\sevensy \scriptscriptfont2=\fivensy
\textfont3=\tenex \scriptfont3=\tenex \scriptscriptfont3=\tenex
\def\rm{\fam0\ninerm}%
\textfont\itfam=\nineit
\def\it{\fam\itfam\nineit}%
\textfont\bffam=\ninebf
\def\bf{\fam\bffam\ninebf}%
\normalbaselineskip=11pt
\setbox\strutbox=\hbox{\vrule height8pt depth3pt width0pt}%
\normalbaselines\rm}

\hsize30cc
\vsize44cc
\nopagenumbers

\def\luz#1{\luzno#1?}
\def\luzno#1{\ifx#1?\let\next=\relax\yyy
\else \let\next=\luzno#1\xxx\fi\next}
\def\sp#1{\def\xxx{\kern1.7pt}\def\yyy{\kern-1.7pt}\luz{#1}}
\def\spa#1{\def\xxx{\kern1pt}\def\yyy{\kern-1pt}\luz{#1}}

\newcount\beg
\newbox\aabox
\newbox\atbox
\newbox\fpbox
\def\abbrevauthors#1{\setbox\aabox=\hbox{\sevenrm\uppercase{#1}}}
\def\abbrevtitle#1{\setbox\atbox=\hbox{\sevenrm\uppercase{#1}}}
\long\def\pag{\beg=\pageno
\def\leftheadline{\noindent\rlap{\nine\folio}\hfil\copy\aabox\hfil}
\def\rightheadline{\noindent\hfill\copy\atbox\hfill\llap{\nine\folio}}
\def\phead{\setbox\fpbox=\hbox{\sevenrm
************************************************}%
\noindent\vbox{\sevenrm\baselineskip9pt\hsize\wd\fpbox%
\centerline{***********************************************}

\centerline{BANACH CENTER PUBLICATIONS, VOLUME **}

\centerline{INSTITUTE OF MATHEMATICS}

\centerline{POLISH ACADEMY OF SCIENCES}

\centerline{WARSZAWA 19**}}\hfill}
\footline{\ifnum\beg=\pageno \hfill\nine[\folio]\hfill\fi}
\headline{\ifnum\beg=\pageno\phead
\else
\ifodd\pageno\rightheadline \else \leftheadline \fi
\fi}}

\newbox\tbox
\newbox\aubox
\newbox\adbox
\newbox\mathbox

\def\title#1{\setbox\tbox=\hbox{\let\\=\cr
\baselineskip14pt\vbox{\twbf\tabskip 0pt plus15cc
\halign to\hsize{\hfil\ignorespaces \uppercase{##}\hfil\cr#1\cr}}}}

\newbox\abbox
\setbox\abbox=\vbox{\vglue18pt}

\def\author#1{\setbox\aubox=\hbox{\let\\=\cr
\nine\baselineskip12pt\vbox{\tabskip 0pt plus15cc
\halign to\hsize{\hfil\ignorespaces \uppercase{\spa{##}}\hfil\cr#1\cr}}}%
\global\setbox\abbox=\vbox{\unvbox\abbox\box\aubox\vskip8pt}}

\def\address#1{\setbox\adbox=\hbox{\let\\=\cr
\nine\baselineskip12pt\vbox{\it\tabskip 0pt plus15cc
\halign to\hsize{\hfil\ignorespaces {##}\hfil\cr#1\cr}}}%
\global\setbox\abbox=\vbox{\unvbox\abbox\box\adbox\vskip16pt}}

\def\mathclass#1{\setbox\mathbox=\hbox{\footnote{}{1991 {\it Mathematics
Subject
Classification}\/: #1}}}

\long\def\maketitlebcp{\pag\unhbox\mathbox
\footnote{}{The paper is in final form and no version
of it will be published elsewhere.}
\vglue7cc
\box\tbox
\box\abbox
\vskip8pt}

\long\def\abstract#1{{\nine{\bf Abstract.}
#1

}}

\def\section#1{\vskip-\lastskip\vskip12pt plus2pt minus2pt
{\bf #1}}

\long\def\th#1#2#3{\vskip-\lastskip\vskip4pt plus2pt
{\sc #1} #2\hskip-\lastskip\ {\it #3}\vskip-\lastskip\vskip4pt plus2pt}

\long\def\defin#1#2{\vskip-\lastskip\vskip4pt plus2pt
{\sc #1} #2 \vskip-\lastskip\vskip4pt plus2pt}

\long\def\remar#1#2{\vskip-\lastskip\vskip4pt plus2pt
\sp{#1} #2\vskip-\lastskip\vskip4pt plus2pt}

\def\Proof{\vskip-\lastskip\vskip4pt plus2pt
\sp{Proo{f.}\ }\ignorespaces}

\def\endproof{\nobreak\kern5pt\nobreak\vrule height4pt width4pt depth0pt
\vskip4pt plus2pt}

\newbox\refbox
\newdimen\refwidth
\long\def\references#1#2{{\nine
\setbox\refbox=\hbox{\nine[#1]}\refwidth\wd\refbox\advance\refwidth by 12pt%
\def\textindent##1{\indent\llap{##1\hskip12pt}\ignorespaces}
\vskip24pt plus4pt minus4pt
\centerline{\bf References}
\vskip12pt plus2pt minus2pt
\parindent=\refwidth
#2

}}

\def\footnoterule{\kern -3pt \hrule width 4cc \kern 2.6pt}

\catcode`@=11
\def\vfootnote#1%
{\insert\footins\bgroup\nine\interlinepenalty\interfootnotelinepenalty%
\splittopskip\ht\strutbox\splitmaxdepth\dp\strutbox\floatingpenalty\@MM%
\leftskip\z@skip\rightskip\z@skip\spaceskip\z@skip\xspaceskip\z@skip%
\textindent{#1}\footstrut\futurelet\next\fo@t}
\catcode`@=12

\mathclass{Primary 58F05; Secondary 70G50, 58G20.}

\abbrevauthors{V.O. Soloviev}
\abbrevtitle{Divergences and boundary terms}

\title{Divergences in formal variational calculus\\
and boundary terms\\ in Hamiltonian formalism}

\author{Vladimir\ O.\ Soloviev}
\address{Institute for High Energy Physics\\
142 284, Protvino, Moscow region, Russia\\
E-mail: vosoloviev@mx.ihep.su}

\maketitlebcp

\def\d{\mathop{\rm d}}
\def\Rn{\mathop{\rm R^n}}
\def\inprod{\mathop{\kern -0.05em\raise -0.1em\hbox{%
  \vrule height 0.03em width 0.6em depth 0em%
  \vrule height 0.7em width 0.03em depth 0em}\kern 0.1em}\nolimits}

\abstract{It is shown how to extend the formal variational calculus
in order to incorporate integrals of divergences into it. Such a
generalization permits to study nontrivial boundary problems in field
theory on the base of canonical formalism.}

\section{1. Introduction.} The Hamiltonian formulation of classical mechanics
[Arn]
is based on geometrical constructions which use such notions as
differential forms, vector fields and multivectors, and such operations as
differential, interior product, Lie derivative, Schouten-Nijenhuis bracket.
Most of these constructions were extended to field theory in the process of
studying nonlinear integrable models during the last 20 years [Olv86]. This
approach has been called the formal variational calculus [GD]
because it ignores
any terms arising as a result of integration by parts. This is fully
justified in case of periodic boundary conditions or fast decay of
fields at spatial infinity, but
unfortunately, this method is not applicable in its initial form to many
other problems interesting from physical point of view. For example, massless
fields are slowly decaying at infinity and, as a result, some important
characteristics of these fields are expressed just through surface integrals
(or volume integrals of spatial divergences). They are necessary to form
canonical generators of the global gauge transformations or asymptotic
symmetries of the Riemannian metric. The great efforts were started
at the end of fifties to
understand the role of surface terms in the Hamiltonian of
General Relativity [ADM].
Only after 15 years of study the satisfactory explanation had been given [RT].
But even then not all questions were answered. For example, one might worry
how to retain the surface terms which are necessary to realize the Poincar\'e
algebra in asymptotically flat space [RT], [Sol85]
$$
\{H(\xi),H(\eta)\}=H([\xi,\eta])
$$
when doing local calculations of the constraints algebra
$$
\{{\cal H}(x),{\cal H}(y)\}=g^{ab}(x){\cal H}_b(x)\delta_{,a}(x,y)-
g^{ab}(y){\cal H}_b(y)\delta_{,a}(y,x),
$$
$$
\{{\cal H}(x),{\cal H}_a(y)\}=-{\cal H}(y)\delta_{,a}(y,x),
$$
$$
\{{\cal H}_a(x),{\cal H}_b(y)\}={\cal H}_b(x)\delta_{,a}(x,y)-
{\cal H}_a(y)\delta_{,b}(y,x).
$$

We will show in the following that all the main structures of the formal
variational calculus can be extended to include nontrivial contributions
of divergences through introduction of a
new grading and new pairing compatible with it. So, it occurs possible
to preserve the nice geometrical language  in the more general case
than before. After all the field theory Poisson bracket is given by a new
 formula which differs from the standard one by surface terms.
 Simultaneously we get the answer to the mentioned problem of disappearance
 of the surface contributions in local calculations with $\delta$-function.
The natural way to take the boundary terms into account is to introduce
the characteristic function $\theta_{\Omega}(x)$
of the integration domain $\Omega$. Then relations like
$$
\biggl(\theta_{\Omega}(x){{\partial}\over{\partial
x^i}}+\theta_{\Omega}(y){{\partial}\over{\partial y^i}}\biggr) =
-{{\partial\theta_{\Omega}(x)}\over{\partial x^i}}\delta(x,y),
$$
give the solution.
 In its turn this is connected with the observation [Sol92]
that transformations of the type  (for example, transformation to
Ashtekar's variables)
$$
q^A(x)\to q^A(x),\qquad p_A(x)\to p_A(x)+{{\delta F[q]}\over{\delta q^A(x)}},
$$
 in field theory are canonical only up to boundary contributions, because
the standard Euler-Lagrange variational derivatives in general do not
commute [And76], [And78].

 We expect that boundary conditions should be treated in this formalism
 as a kind of constraints put on the initial data, i.e., they should be
 added to the Hamiltonian with some Lagrange multipliers and then
 checked for compatibility with the dynamics. The requirement of
 compatibility may lead to secondary boundary conditions or to fixing
 the Lagrange multipliers. But now this subject is not enough studied and
 our consideration is preliminary and limited to one example:
the nonlinear Schr\"{o}dinger equation.

\section{2. New Poisson bracket formula.}
Below we  use the local coordinate language and
instead of the manifold with a boundary  consider a domain $\Omega$
in  $\Rn$ having a smooth boundary $\partial\Omega$.
We do not expect that global formulation could meet with serious difficulties.

\defin{Definition}{1.
An integral over  a finite domain $\Omega$
of  a function  of field variables $\phi^A(x), A=1,...,p$
and their partial derivatives $D_J\phi^A$ up to some finite order
$$
F=\int_{\Omega}d^nxf(\phi_A(x),D_J\phi_A(x))
$$
is called a {\it local functional.}\/}

\remar{Remark\ {1.}\ }{In contrast to the standard definition we do not
treat these local
functionals as equivalent if they differ by a divergence term.}

All the functions $f$ and $\phi_A$ as well as their variations
throughout the paper are supposed  infinitely smooth, i.e. $C^{\infty}(\Rn)$.
We  use multi-index notations $J=(j_1,...,j_n)$
$$
D_J={{\partial^{|J|}}\over{\partial^{j_1}x^1...{\partial^{j_n}x^n}}},
\qquad |J|=j_1+...+j_n, \qquad D_0=1.
$$
The derivative operator $D$ will denote below the full partial derivative
taking into account also coordinate dependence of fields $\phi_A(x)$.
As the number of sums in some formulae of this paper is large enough
we will write only a sign of summing without displaying
the indices of summation. According to this rule, sum over all
repeated indices should be understood. In those cases, where it is
not so, we display the summation indices. Also, we do not show the limits
of summation, because they are  natural, i.e. outside them the
summand is simply zero.
Usually we omit $d^nx$ in the integrals
and show the arguments only when they can be mixed.

We denote  as ${\cal  A}$ the space of local functionals . It is
important that this space includes functionals with integrands
depending on derivatives of arbitrary order [And92]. Otherwise the Poisson
brackets could go out of ${\cal  A}$. The following is the general definition
of field theory Poisson bracket.

\defin{Definition}{2
A bilinear operation $\{ \cdot,\cdot\}$   such that
for any $F,G,H\in~{\cal A}$

{\rm 1)} $\{F,G\}\in {\cal A}$;

{\rm 2)} $\{F,G\}=-\{G,F\}$;

{\rm 3)} $\{\{F,G\},H\}+\{\{H,F\},G\} +\{\{G,H\},F\}=0$;

is called the {\it field theory Poisson bracket.}}

The key idea of the new formula is in exploitation of the full variations
which are free on the boundary.
The variation of a local functional
$$
\delta F=\sum\int{{\partial f}\over{\partial\phi_A^{(J)}}}D_J\delta\phi_A
$$
contains the differential operator which is called Fr\'echet derivative
$$
f'_A=\sum_{J=0}^{\infty}{{\partial f}\over {\partial\phi_A^{(J)}}}D_J.
$$
We propose to define field theory Poisson bracket by the formula
$$
\{ F,G \} =
\sum\int_{\Omega} \mathop{\rm Tr}(D_{f_A}\hat I_{AB}D_{g_B}),
$$
where trace of two differential operators is used
$$
\mathop{\rm Tr}(\hat A\hat B)=\sum_{I,K} D_IA_KD_KB_I,
$$
where
$$
\hat A=\sum_K A_KD_K,\qquad \hat B=\sum_I B_ID_I.
$$
The important property of the trace is
$$
\mathop{\rm Tr}(\hat AD\hat B)=D\mathop{\rm Tr}(\hat A\hat B)=
\mathop{\rm Tr}(D\hat A\hat B).
$$
The individual structure of a Poisson bracket is given by matrix $I_{AB}$.
More general treatment of it will be given in the next Section and
here it is simply constant antisymmetric matrix.

Symmetrized covariant
derivatives can also be used in the expression for the first variation of
local functional . We can replace partial derivatives by covariant ones in
the trace calculation if the curvature is zero or if one operator is
simply a multiplication by a
function. Then covariance of the new formula under changes of independent
variables is evident. In general case a special consideration is necessary.

The new bracket differs from the standard one in the exact fulfilment of
the Jacobi identity under arbitrary boundary conditions [Sol93].
In the same time its calculation is not more complicated than usual
since we need not integrate by parts to get Euler-Lagrange derivatives.

We can also use another representation of the first variation [Olv86]
$$
\delta F=\sum\int_{\Omega} D_J\biggl( E^J_A(f)\delta\phi_A\biggr).
$$
where the higher Eulerian operators [KMGZ], [Ald]
$$
E^J_A(f)=\sum_K (-1)^{|K|+|J|}{K\choose J}D_{K-J}{{\partial f}
\over{\partial\phi_A^{(K)}}},
$$
are used. The zero order operator
is just the standard Euler-Lagrange variational derivative.
Binomial coefficients for multi-indices are
$$
{J \choose K}={j_1\choose k_1}...{j_n\choose k_n},
$$
where  ordinary binomial coefficients are
$$
{j \choose k}= \cases {
j!/(k!(j-k)!) &  if $ 0\le k \le j$; \cr
0  & otherwise. \cr }
$$
Let us mention that if $J$ is not contained in $K$, then all quantities
having multi-index $(K-J)$ are zero. The sums over $J$
and $K$ above are really finite because local functional can
depend only on a finite number of derivatives according to Definition 1.

A remarkable property of these operators is
$$
E^J_A(D_I f)=E_A^{J-I}(f).
$$
It leads to the result that if
$$
\{\int f,\int g\}=\int h,
$$
then
$$
\{\int Df,\int g\}=\{\int f,\int Dg\}=\int Dh.
$$

\section{3. Extension of the formal variational calculus.}
In dealing with terms arising in the integration by parts it is suitable
to represent integrals over finite domain as integrals over infinite space
with the help of the characteristic function
$$
\theta_{\Omega}(x)= \cases {
1 &  if $x\in\Omega$; \cr
0  & otherwise. \cr }
$$
We can understand it also as Heaviside function
$\theta_{\Omega}(x)
=\theta (P_{\Omega})$,
where equation $P_{\Omega}(x)=0$ defines the boundary and
$$
P_{\Omega}(x)\cases {
>0 &  if $x\in\Omega$; \cr
<0  & otherwise. \cr }
$$
Then
$$
F=\int\limits_{\Omega}f=\int\theta_{\Omega}f,
$$
and, for example, the full variation of a local functional can be expressed
in the form
$$
\delta F=\int{{\delta F}\over{\delta\phi_A}}\delta\phi_A,
$$
where  the distribution
$$
{{\delta F}\over{\delta\phi_A}}=E^0_A(\theta_{\Omega}f)=
\sum (-1)^{\vert J\vert}D_J\theta_{\Omega}
E^J_A(f)
$$
could be called the {\it full variational derivative}\/.
Such a representation
corresponds to the situation opposite to the standard one: here
distributions
are of finite support whereas test functions $\delta\phi_A$ are arbitrary.

A {\it  grading}\/ in linear space $L$ is a decomposition of it into
direct sum
of subspaces, with a special value of some function $p$ (grading function)
assigned to all the elements of any subspace [Dorf]. Elements of each
subspace
are called {\it homogeneous}\/.

In our case the factor $D_J\theta_{\Omega}$ is responsible
for the grading
and the function $p$ takes its values in the set of all positive
multi-indices $J=(j_1,\dots,j_n)$
$$
L=\bigoplus\limits_{J=0}^{\infty} L^{\langle J\rangle}.
$$
We always can return to the standard formal variational calculus by
putting $\theta_{\Omega}(x)\equiv 1$.

A bilinear operation $x,y\mapsto x\circ y$, defined on $L$, is said to be
{\it compatible with the grading}\/ if the product of any homogeneous elements
is also homogeneous, and if
$$
p(x\circ y)=p(x)+p(y).
$$

\vskip 4pt plus 2pt
{\bf 3.1.}{\it Local functionals and evolutionary vector fields.}
Here we will call the
expression given in Definition 1 the {\it canonical form of a local
functional}\/. We formally extend
that definition by allowing local functionals to be written as follows
$$
F=\sum_{J=0}^{\infty}\int D_J\theta_{\Omega}(x)f^{\langle J\rangle}
\bigl(\phi_A(x),
D_K\phi_A(x)\bigr)d^nx=\sum\int\theta^{(J)}f^{\langle J\rangle},
$$
where in accordance with the previous definition only a finite number
of terms is allowed. Here and below we simplify the notation
for derivatives of $\theta$ and remove $\Omega$.
Of course, any such functional can be transformed
to the form used above through integration by parts
with
$$
f=\sum(-1)^{|J|}D_Jf^{\langle J\rangle}.
$$
So, the formal integration by parts over infinite
space $\Rn$ evidently changes the grading.
It will be clear below that the general
situation is from one side compatibility of all bilinear operations
with the grading and from the other side compatibility of them
with formal integration by parts.
So, basic objects (local functionals etc.) are defined as equivalence
classes modulo formal divergences (i.e., divergences of expressions
containing $\theta$-factors) and the unique
decomposition into homogeneous subspaces with fixed grading function
can be made only for representatives of these classes. But we will see
that the pairing will be defined in such a way to avoid any ambiguity.

We call  expressions of the form
$$
\psi=\sum\int\theta^{(J)}
D_K\psi^{\langle J\rangle}_A{{\partial}\over{\partial\phi_A^{(K)}}}
$$
the {\it evolutionary vector fields}\/. The expressions
$\psi^{\langle J\rangle}_A$ are called {\it characteristics} of them.
The value of the evolutionary vector field on a local functional is given
by formula
$$
\psi F=\sum\int\theta^{(I+J)}
D_K\psi^{\langle J\rangle}_A{{\partial f^{\langle I
\rangle}}\over{\partial\phi_A^{(K)}}}.
$$
It is a straightforward calculation to check that this operation is
compatible with the formal integration by parts, i.e.
$$
\psi{\rm Div}(f)={\rm Div}(\psi f),
$$
similarly to in the standard formal variational calculus. This relation is,
of course, valid for integrands.

It is easy to check that the evolutionary vector field with coefficients
$$
\psi^{(J)}_A=\sum
\biggl( D_L\xi_B^{\langle I\rangle} {{\partial\eta_A^{\langle J-I\rangle
}}\over{\partial \phi_B^{(L)}}}-
D_L\eta_B^{\langle I\rangle} {{\partial\xi_A^{\langle J-I\rangle}}\over
{\partial \phi_B^{(L)}}}\biggr)
$$
can be considered as the {\it  commutator of the evolutionary vector fields}
 $\xi$ and $\eta$
$$
\psi F=[\xi,\eta]F=\xi(\eta F)-\eta(\xi F),
$$
with the Jacobi identity fulfilled for the commutator operation.
Therefore the vector fields form a Lie algebra.

\vskip 4pt plus 2pt
{\bf 3.2.}{\it Differentials and functional forms.}
The {\it differential of a local functional}\/
is simply the first variation of it
$$
\d F=\sum\int\theta^{(J)}
{{\partial f^{\langle J\rangle}}\over{\partial\phi_A^{(K)}}}\delta\phi_A^{(K)},
$$
here and below $\delta\phi_A^{(K)}=D_K\delta\phi_A$.
It can also be expressed through
the Fr\'echet derivative
or through the higher Eulerian operators
$$
\d F=\sum\int\theta^{(J)}{f^{\langle J\rangle}}'(\delta\phi)=
\sum\int\theta^{(J)}
D_K\bigl( E^K_A(f^{\langle J\rangle})\delta\phi_A\bigr) .
$$
This differential is a special example of functional 1-form.
A general functional 1-form can be written as
$$
\alpha = \sum\int\theta^{(J)}\alpha ^{\langle J\rangle}_{AK}
\delta\phi_A^{(K)}.
$$
Of course, the coefficients are not unique since we can do
formal integration by parts.

Let us call the following
expression the {\it canonical form of functional 1-form}
$$
\alpha=\sum\int\theta^{(J)}\alpha^{\langle J\rangle}_A\delta\phi_A.
$$

Analogously, we can define {\it functional $m$-forms}
as integrals, or equivalence classes modulo formal
divergences, of vertical forms
$$
\alpha ={{1}\over{m!}}
\sum\int\theta^{(J)}\alpha^{\langle J\rangle}_{A_1K_1,\dots,A_mK_m}\delta
\phi_{A_1}^{(K_1)}\wedge\dots\wedge\delta\phi_{A_m}^{(K_m)}.
$$

Define the {\it pairing}\/
of an evolutionary vector field and 1-form as
$$
\alpha (\xi)=\xi\inprod\alpha=\sum
\int\theta^{(I+J)}\alpha ^{\langle J\rangle}_{AK}D_K\xi_A^{\langle I\rangle}.
\leqno(1)
$$

The {\it interior product}\/
of an evolutionary vector field and functional $m$-form
will be
$$
\xi\inprod\alpha=
{{1}\over{m!}}\sum(-1)^{i+1}
\int\theta^{(J+I)}\alpha^{\langle J\rangle}_{A_1K_1,\dots,A_mK_m}
D_{K_i}\xi_{A_i}^{\langle I\rangle}\delta
\phi_{A_1}^{(K_1)}\wedge\dots
$$
$$
\dots\wedge\delta\phi_{A_{i-1}}^{(K_{i-1})}
\wedge\delta\phi_{A_{i+1}}^{(K_{i+1})}\wedge\dots\wedge
\delta\phi_{A_m}^{(K_m)}.
$$
Then the value of $m$-form on the $m$ evolutionary vector fields
will be defined by  formula
$$
\alpha (\xi_1,\dots,\xi_m)=\xi_m\inprod\dots\xi_1\inprod\alpha.
$$
It can be  checked by straightforward calculation that
$$
{\rm Div}(\alpha) (\xi_1,\dots,\xi_m)=
{\rm Div}(\alpha (\xi_1,\dots,\xi_m)).
$$

The {\it differential of  $m$-form} given as
$$
\d\alpha ={{1}\over{m!}}
\sum
\int\theta^{(J)}{{\partial\alpha^{\langle J
\rangle}_{A_1K_1,\dots,A_mK_m}}\over
{\partial\phi_A^{(K)}}}\delta\phi_A^{(K)}\wedge\delta
\phi_{A_1}^{(K_1)}\wedge\dots
\wedge\delta\phi_{A_m}^{(K_m)},
$$
satisfies standard properties
$$
{\d}^2=0
$$
and
$$
\d\alpha(\xi_1,\dots,\xi_{m+1})=
\sum\limits_i(-1)^{i+1}\xi_i\alpha(\xi_1,\dots,
\hat\xi_i,\dots,\xi_{m+1})+
$$
$$
+\sum\limits_{i<j}(-1)^{i+j}\alpha([\xi_i,\xi_j],\xi_1,\dots,\hat\xi_i,\dots,
\hat\xi_j,\dots,\xi_{m+1}).
$$

The {\it Lie derivative}\/ of a functional form $\alpha$
along an evolutionary vector field $\xi$ can be introduced by the standard
formula
$$
L_{\xi}\alpha=\xi\inprod\d\alpha+\d (\xi\inprod\alpha).
$$

\vskip 4pt plus 2pt
{\bf 3.3.}{\it Graded differential operators and their adjoints.}\/
We call linear differential operators of the form
$$
\hat I=\sum_{J=0}^{\infty}\theta^{(J)}
\sum_{N=0}^{N_{max}}I^{\langle J\rangle N}_{AB}D_N
$$
{\it graded differential operators}\/.

Let us call linear differential operator $\hat I^{\ast}$ {\it adjoint}\/ to
$\hat I$ if for arbitrary
set of smooth functions $f_A$, $g_A$
$$
\sum\limits_{A,B}\int f_A\hat I_{AB}g_B=
\sum\limits_{A,B}\int g_A\hat I^{\ast}_{AB}f_B.
$$
For coefficients of the adjoint operator we can derive the expression
$$
I^{\ast\langle J\rangle M}_{AB}=\sum\limits_{K=0}^{K_{max}}
\sum\limits_{L=0}^{min(K,J)}
(-1)^{|K|}{K\choose L}{K-L\choose M}
D_{K-L-M}I^{\langle J-L\rangle K}_{BA}.
\leqno(2)
$$
It is easy to check that the relation
$$
\hat I(x)\delta(x,y)=\hat I^{\ast}(y)\delta(x,y)
$$
is valid. For example, we have
$$
\biggl(\theta(x){{\partial}\over{\partial
x^i}}+\theta(y){{\partial}\over{\partial y^i}}\biggr) =-\theta^{(i)}\delta
(x,y).
$$

Operators satisfying relation
$$
\hat I^{\ast}=-\hat I
$$
will be called {\it skew-adjoint}\/. With the help of them it is possible to
express 2-forms (and also 2-vectors to be defined below) in the
canonical form
$$
\alpha={{1}\over{2}}\sum\limits_{A,B}\int\delta\phi_A\wedge\hat I_{AB}
\delta\phi_B.
$$
It is clear that we can consider these representations of functional forms
as formal decompositions over the basis derived as result of the tensor product
of $\delta\phi_A$, with the totally antisymmetric multilinear operators
$$
\hat\alpha=\sum\theta^{(J)}\alpha^{\langle J\rangle}_{A_1K_1,\dots,A_mK_m}
\biggl( D_{K_1}\cdot,\dots,D_{K_m}\cdot\biggr)
$$
being coefficients of these decompositions.

\vskip 4pt plus 2pt
{\bf 3.4.}{\it Multi-vectors and Schouten-Nijenhuis bracket.}\/
Let us introduce the dual basis to $\vert\delta\phi_A\rangle$ by
formal relation
$$
\left\langle{{\delta}\over{\delta\phi_B(y)}},\delta\phi_A(x)\right\rangle
=\delta_{AB}\delta(x,y),
$$
and construct by means of the tensor product a basis
$$
{{\delta}\over{\delta\phi_{B_1}(y)}}\otimes{{\delta}\over{\delta\phi_{B_2}(y)}}
\otimes\dots\otimes{{\delta}\over{\delta\phi_{B_m}(y)}}.
$$

Then by using totally antisymmetric
multilinear operators we can
define {\it functional $m$-vectors}\/ (or {\it multi-vectors}\/)
$$
\psi={{1}\over{m!}}\sum\int\theta^{(J)}
\psi^{\langle J\rangle}_{B_1L_1,\dots,B_mL_m}D_{L_1}
{{\delta}\over{\delta\phi_{B_1}}}\wedge\dots\wedge D_{L_m}
{{\delta}\over{\delta\phi_{B_m}}}.
$$
Here a natural question arises: what is the relation between evolutionary
vector fields and 1-vectors?
Evidently, evolutionary vector fields lose their form when
integrated by parts whereas 1-vectors conserve it.
It is possible to prove the following Proposition [Sol94].

\th{Proposition}{1.}
{There is a one-to-one correspondence between evolutionary vector
fields and functional 1-vectors.
The  coefficients of 1-vector in the canonical form $\xi_A^{\langle J
\rangle}$ are equal to the characteristic
of the evolutionary vector field.}

It is not difficult to show that we can  define pairing (interior product) of
1-forms and 1-vectors and this pairing preserves the identification
$$
\alpha (\xi)=
\sum\int\theta^{(I+J)}\mathop{\rm Tr}(\alpha^{\langle I\rangle}
\xi^{\langle J\rangle}).
$$
When 1-vector is in the canonical form this result coincides with Eq.(1).

The interior product of 1-vector and $m$-form or, analogously,
of 1-form and $m$-vector is defined as
$$
\xi\inprod\alpha={{1}\over{m!}}\sum (-1)^{(i+1)}\int
\theta^{(I+J)}D_{K_i}\xi^{\langle I\rangle}_{A_iL}D_L
\biggl(\alpha^{\langle J\rangle}_{A_1K_1,\dots,
A_mK_m}\delta\phi_{A_1}^{(K_1)}\wedge\dots
$$
$$
\dots\wedge\delta\phi_{A_{i-1}}^{(K_{i-1})}\wedge
\delta\phi_{A_{i+1}}^{(K_{i+1})}\wedge\dots
\wedge\delta\phi_{A_m}^{(K_m)}\biggr).
$$
Then we also can define the value of $m$-form on $m$ 1-vectors (or,
analogously, $m$-vector on $m$ 1-forms)
$$
\alpha (\xi_1,\dots,\xi_m)=
\xi_m\inprod\dots\xi_1\inprod\alpha=
\sum\int\theta^{(J+I_1+\dots+I_m)}\mathop{\rm Tr}
\biggl( \alpha^{\langle J\rangle}
\xi_1^{\langle I_1\rangle}\cdots\xi_m^{\langle I_m\rangle}\biggr),
$$
where in this trace
each entry of multilinear operator $\alpha$ acts only to the one
corresponding $\xi$, whereas each derivation of the operator $\xi$ acts on the
product of $\alpha$ and all the rest of $\xi$'s.

It is possible to extend the differential onto $m$-vectors
$$
\d\psi ={{1}\over{m!}}\sum\int\theta^{(J)}
{{\partial\psi^{\langle J\rangle}_
{A_1K_1,\dots,A_mK_m}}\over{\partial\phi_B^{(L)}}}\delta\phi_B^{(L)}D_{K_1}
{{\delta}\over{\delta\phi_{A_1}}}\wedge\dots\wedge
D_{K_m}{{\delta}\over{\delta\phi_{A_m}}},
$$
and analogously onto mixed  objects. Evidently, ${\rm d}^2\psi=0$.

With the help of the previous constructions we can define the
{\it Schouten-Nijenhuis bracket} as follows
$$
\bigl[ \xi,\eta\bigr]_{SN} =\d\xi\inprod\eta + (-1)^{pq}\d
\eta\inprod\xi
$$
for two multi-vectors of orders $p$ and $q$. The result of this operation is
$p+q-1$-vector and it is analogous to the Schouten-Nijenhuis bracket in
tensor analysis [Nij]. Its use in formal variational calculus is
described in [Dorf]. However
this bracket is defined only for operators there.
We can recommend [Olv84] as
an interesting source for treatment of the Schouten-Nijenhuis bracket
for functional multi-vectors.
Our construction of this bracket
guarantees compatibility with the equivalence modulo divergences
$$
\bigl[ {\rm Div}(\xi),\eta\bigr]_{SN} ={\rm Div}\bigl[ \xi,\eta\bigr]_{SN}=
\bigl[ \xi, {\rm Div}(\eta)\bigr]_{SN}.
$$

\th{Proposition}{2.}
{
The Schouten-Nijenhuis bracket of functional 1-vectors up to a sign
coincides with the commutator of corresponding evolutionary vector
fields.}

\Proof
Let us take two 1-vectors in canonical form
$$
\xi=\sum\int\theta^{(J)}\xi^{\langle J\rangle}_A
{{\delta}\over{\delta\phi_A}},\qquad
\eta=\sum\int\theta^{(K)}\eta^{\langle K\rangle}_B{{\delta}
\over{\delta\phi_B}},
$$
and compute
$$
\bigl[ \xi,\eta\bigr]_{SN}=\d\xi\inprod\eta -\d\eta\inprod\xi.
$$
We have
$$
\d\xi=\sum\int\theta^{(J)}{\xi^{\langle J\rangle}_A}'(\delta\phi){{\delta}
\over{\delta
\phi_A}}=\sum\int\theta^{(J)}{{\partial\xi^{\langle J\rangle}_A}\over
{\partial\phi^{(L)}_C}}\delta\phi_C^{(L)}{{\delta}\over{\delta\phi_A}},
$$
and
$$
\d\xi\inprod\eta=-\sum\int\theta^{(J+K)}{{\partial
\xi_A^{\langle J\rangle}}\over
{\partial\phi_B^{(L)}}}D_L\eta_B^{\langle K\rangle}{{\delta}\over
{\delta\phi_A}}.
$$
Therefore, we obtain
$$
\bigl[ \xi,\eta\bigr]_{SN}=-\sum\int\theta^{(J+K)}\biggl(
D_L\eta_B^{\langle K\rangle}{{\partial\xi_A^{\langle J\rangle}}\over
{\partial\phi_B^{(L)}}}-
D_L\xi_B^{\langle K\rangle}{{\partial\eta_A^{\langle J\rangle}}\over
{\partial\phi_B^{(L)}}}
\biggr)
{{\delta}\over{\delta\phi_A}}=-[\xi,\eta].
$$

\vskip 4pt plus 2pt

\th{Proposition }{3. (Olver's Lemma [Olv86])}
{
The Schouten-Nijenhuis bracket of the two bivectors can be expressed
in the form
$$
\bigl[ \xi,\psi\bigr]_{SN}=-{{1}\over{2}}\sum\int\
\xi\wedge\hat I'(\hat K\xi)\wedge\xi
-{{1}\over{2}}\sum\int\
\xi\wedge\hat K'(\hat I\xi)\wedge\xi,
$$
where the two differential operators $\hat I$, $\hat K$ are the
coefficients of the bivectors in their canonical form.}

\Proof
Let us consider the Schouten-Nijenhuis bracket for the two bivectors and
without loss of generality
take them in the canonical form
$$
\chi={{1}\over{2}}\sum\int\theta^{(L)}\xi_A\wedge I^{\langle L\rangle N}_{AB}
D_N\xi_B,
$$
$$
\psi={{1}\over{2}}\sum\int\theta^{(M)}\xi_C\wedge K^{\langle M\rangle P}_{CD}
D_P\xi_D,
$$
where $\xi_A={\delta}/{\delta\phi_A}$ and operators $\hat I$ , $\hat K$
are skew-adjoint. Then we have
$$
\d\chi={{1}\over{2}}\sum\int\theta^{(L)}{{\partial
I^{\langle L\rangle N}_{AB}}\over
{\partial\phi_E^{(J)}}}\delta\phi_E^{(J)}\xi_A\wedge D_N\xi_B,
$$
and
$$
\d\chi\inprod\psi={{1}\over{4}}\sum\int\theta^{(L+M)}
{{\partial I^{\langle L\rangle N}_{AB}}\over{\partial\phi_C^{(J)}}}D_J
\biggl( K^{\langle M\rangle P}_{CD}D_P\xi_D\biggr)\wedge
\xi_A\wedge D_N\xi_B-
$$
$$
-{{1}\over{4}}\sum\int\theta^{(L+M)}D_P\biggl(
{{\partial I^{\langle L\rangle N}_{AB}}\over
{\partial\phi_D^{(J)}}}\xi_A\wedge D_N\xi_B\biggr)
\wedge D_J(\xi_C K^{\langle M\rangle P}_{CD}).
$$
Now let us make integration by parts in the second term
$$
\d\chi\inprod\psi=-{{1}\over{4}}\sum\int\theta^{(L+M)}\xi_A\wedge
(I^{\langle L\rangle N}_{AB})'\biggl(\hat K^{\langle M\rangle}
\xi\biggr)\wedge  D_N\xi_B-
$$
$$
-{{1}\over{4}}\sum\int\theta^{(L+M+Q)}(-1)^{|P|}{P\choose Q}
{{\partial I^{\langle L\rangle N}_{AB}}\over
{\partial\phi_D^{(J)}}}\xi_A\wedge D_N\xi_B
\wedge D_{J+P-Q}(\xi_C K^{\langle M\rangle P}_{CD}).
$$
At last we change the order of multipliers under wedge product
in the second term,
make a replacement $M\rightarrow M-Q$ and organize the
whole expression in the form
$$
\d\chi\inprod\psi=-{{1}\over{4}}\sum\int\theta^{(L+M)}\xi_A\wedge
(I^{\langle L\rangle N}_{AB})'_C\Biggl(\hat K^{\langle M\rangle}_{CD}\xi_D+
$$
$$
+(-1)^{|P|}{P\choose Q}{P-Q\choose R}
D_{P-Q-R}K^{\langle M-Q\rangle P}_{CD}D_R\xi_C \Biggr)\wedge D_N\xi_B.
$$
Having in mind the definition of adjoint operator (2) we can
represent the final result of the calculation as follows,
$$
\bigl[ \xi,\psi\bigr]_{SN}=-{{1}\over{2}}\sum\int\theta^{(L+M)}
\xi\wedge\biggl((\hat I^{\langle L\rangle})'(\hat K^{\langle M\rangle}\xi)
+(\hat K^{\langle M\rangle})'(\hat I^{\langle L\rangle}\xi)\biggr)\wedge\xi,
$$
therefore supporting in this extended formulation the method, proposed
in [Olv86] for testing the Jacobi identity.

\vskip 4pt plus 2pt

{\bf 3.5.}{\it Poisson brackets and Hamiltonian vector fields.}\/
Let us call a bivector
$$
\Psi={{1}\over{2}}\sum\int{{\delta}\over{\delta\phi_A}}\wedge\hat I_{AB}
{{\delta}\over{\delta\phi_B}},
$$
formed with the help of the graded skew-adjoint differential operator
$$
\hat I_{AB}=\sum \theta^{(L)}I^{\langle L\rangle N}_{AB}D_N,
$$
the {\it Poisson bivector}\/ if
$$
\bigl[ \Psi,\Psi\bigr]_{SN} =0.
$$

The operator $\hat I_{AB}$ is then called the {\it Hamiltonian operator}.

We may call the value of the Poisson bivector on differentials of
the two functionals $F, G$
$$
\{ F,G\} = \Psi (\d F,\d G)=\d G\inprod\d F\inprod\Psi
$$
the {\it Poisson bracket}\/ of these functionals.

The explicit form of the Poisson brackets can easily be obtained. It
depends on the explicit form of the differential of the functionals,
which can be changed by partial integration. Of course, all the
possible forms are equivalent. Taking the extreme cases we have the
expression through Fr\'echet derivatives
$$
\{ F,G \} =
\sum\int\theta^{(J)} \mathop{\rm Tr}\bigl( f'_A\hat I^{\langle J\rangle}_{AB}
g'_B \bigr),
\leqno(3)
$$
or through higher Eulerian operators
$$
\{ F,G \} =
\sum\int\theta^{(J)} D_{P+Q}\bigl( E^P_A(f)\hat I^{\langle J\rangle}_{AB}
E^Q_B(g)
\bigr).
\leqno(4)
$$

\th{Theorem}{1.}
{The Poisson bracket defined above satisfy Definition 1.}

\Proof
It follows from the three
facts: 1)from the previous formulas (3), (4)
it is clear that $\{ F,G \}$ is a local functional,
2)antisymmetry of $\{ F,G \}$ as a consequence of skew-adjointness
of $\hat I_{AB}$ and
3)equivalence of the Jacobi identity
to the Poisson bivector property can be proved [Sol94].

\vskip 4pt plus 2pt

The result of interior product of the differential of a local
functional $H$ and the Poisson bivector (up to the sign)
will be called the {\it  Hamiltonian
vector field}\/ (or the {\it Hamiltonian 1-vector}\/)
$$
\hat I\d H=-\d H\inprod\Psi
$$
corresponding to the Hamiltonian $H$.

Evidently, the standard relations take place
$$
\{ F,H\} = \d F(\hat I \d H)=(\hat I \d H)F.
$$

\th{Theorem}{2.}
{The Hamiltonian vector field corresponding to  Poisson bracket
of the functionals $F$ and $H$ coincides up to the sign with
commutator of the Hamiltonian vector fields corresponding to these
functionals.}

\Proof
Consider the value of commutator of Hamiltonian vector fields
$\hat I\d F$ and $\hat I\d H$ on arbitrary functional $G$
$$
[\hat I\d F, \hat I\d H]G=\hat I\d F(\hat I\d H(G))-\hat I\d H(\hat I\d F(G))
=
$$
$$
=\hat I\d F(\{G,H\})-\hat I\d H(\{G,F\})=\{\{G,H\},F\}-\{\{G,F\},H\}=
$$
$$
=-\{G,\{F,H\}\}=-\hat I\d\{F,H\}(G),
$$
where we have used the Jacobi identity and antisymmetry of Poisson bracket.
Due to arbitrariness of $G$ the proof is completed.

\vskip 4pt plus 2pt

\remar{Example\ {1.}\ }{
Let us consider a first structure
$$
\{ u(x),u(y)\} =D_x\delta(x,y)
$$
of the Korteweg-de Vries equation ([Olv86] Example 7.6)
$$
u_t=u_{xxx} +uu_x.
$$
Construct the adjoint graded operator
to $\theta D$ according to Eq.(2)
$$
(\theta D)^{\ast}=-\theta D -D\theta,
$$
and the skew-adjoint operator is
$$
\hat I={{1}\over{2}}\biggl(\theta D- (\theta D)^{\ast}\biggr)=\theta D +
{{1}\over{2}}D\theta.
$$
The Poisson bivector has the form
$$
\Psi={{1}\over{2}}\int\theta\biggl( {{\delta}\over{\delta u}}\wedge
D{{\delta}\over{\delta u}}\biggr).
$$
Differential of a local functional $H$ (for simplicity we
suppose it is written in canonical
$$
H=\int\theta h
$$
form) is equal to
$$
\d H=\int\theta h'(\delta u)=\sum_{k=0}^{\infty}
\int\theta^{(k)}(-1)^kE^k(h)\delta u,
$$
where Fr\'echet derivative or higher Eulerian operators can be used.
Therefore, the Hamiltonian vector field generated by $H$ is
$$
\hat I\d H=-\d H\inprod\Psi=
-{{1}\over{2}}\int\theta\biggl[ h'\bigl( D{{\delta}\over{\delta u}}
\bigr) - Dh'\bigl( {{\delta}\over{\delta u}}\bigr)\biggr] ,
$$
or
$$
-{{1}\over{2}}\int\theta^{(k)}(-1)^k\biggl[ E^k(h)D-
DE^k(h)\biggr]{{\delta}\over{\delta u}},
$$
or also
$$
-{{1}\over{2}}\int\theta^{(k)}(-1)^kD_i\biggl[ E^k(h)D-
DE^k(h)\biggr]{{\partial}\over{\partial u^{(i)}}}.
$$
The value of this vector field on another functional $F$ coincides with
the Poisson bracket
$$
-\d F\inprod\d H\inprod\Psi=\{ F,H\}=
{{1}\over{2}}\sum\int\limits_{\Omega}D_{k+l}
\biggl( E^k(f)DE^l(h)-E^k(h)DE^l(f)\biggr).
$$
}

\section{4. Dynamics on a boundary: an example.}
Let us now consider a simple Hamiltonian system in order to obtain
boundary equations with the help of the new brackets.
The nonlinear Schr\"{o}dinger equation
$$
i\dot\psi=-\psi''+2k\psi\vert\psi\vert^2,
$$
can be treated [LR] as generated by the Hamiltonian
$$
H={{1}\over{2}}\int({\cal H}+\bar{\cal H})dx,
$$
where
$$
{\cal H}=r' q' +kr^2q^2, \qquad \bar{\cal H}=\bar{r'}\bar{q'}+
k{\bar{r}}^2{\bar{q}}^2,
$$
and Poisson brackets are
$$
\{q(x),r(y)\}=-2i\delta (x,y), \qquad\{\bar{q}(x),\bar{r}(y)\}=2i\delta (x,y).
$$
To return to the standard form of this equation
we should put reality conditions
$$
\psi=q=\bar{r}, \qquad\bar\psi=r=\bar{q}.
$$
Let us calculate the full variational derivatives
$$
{{\delta H}\over{\delta q}}=kr^2q-{1\over 2}r''-{1\over 2}\theta'r',
\qquad
{{\delta H}\over{\delta r}}=krq^2-{1\over 2}q''-{1\over 2}\theta'q',
$$
analogous formulas take place for the bar variables.

The  natural boundary condition arises if we put
$\delta$-function contribution on the boundary to zero by taking
$$
q'=r'=\bar q'=\bar r'=0.
$$
By considering Poisson brackets for integrals of the total spatial
derivatives of canonical variables $\phi_A=(q,r)$ with the Hamiltonian
we expect to obtain dynamical equations on the boundary in functional
form
$$
{{d}\over{dt}}\int{\phi_A}' dx=
\{\int{\phi_A}' dx, H\}=\int D\bigl( {{1}\over{2}}\sum\limits_BI_{AB}
{{\partial\cal H}\over{\partial\phi_B}}\bigr) dx.
$$
Using Newton-Leibnitz formula we get
$$
\dot{\phi_A}\bigg\vert_1^2={{1}\over{2}}\sum\limits_BI_{AB}
{{\partial\cal H}\over{\partial\phi_B}}\bigg\vert_1^2.
$$
This is different from the standard (bulk) equations
$$
\dot{\phi_A}={{1}\over{2}}\sum\limits_BI_{AB}E^0_B(\cal H).
$$
For the given Hamiltonian the formal equations for boundary values are
$$
\dot q_b=-2ikr_b{q_b}^2, \qquad\dot r_b=2ik{r_b}^2q_b,
$$
if we assume independent behaviour at the ends.
It is remarkable that the boundary equations are different from
the bulk ones despite the boundary condition.
These equations can be easily integrated and give elementary oscillations
at the ends
$$
\psi_{b}(t)=\psi_{b}(0)exp(-2ik|\psi_{b}(0)|^2t),
$$
where the initial value of $\psi$ determines both amplitude and frequency
of the oscillator.

In this case the dynamics of boundary values is separated
from the bulk dynamics. Of course, this situation is not general.

Let us mention that our boundary condition is compatible with the dynamics,
for example,
$$
{d\over{dt}}q'\vert_1^2=\{\int q'',H\}=-2ikr'\vert_1^2=0.
$$

\section{5. Discussion.}
There are not so many publications on problems
where divergences play a nontrivial role
in Hamiltonian formalism of field theory.
After the classical paper by Regge and Teitelboim [RT] we can recommend the
work by Jezierski and Kijowski [JK] (see also book[KT])
where the main criterion is also the
disappearance of surface terms in the first variation of the Hamiltonian.
Such functionals are called admissible or differentiable, but
to be convinced in the full consistency of the formalism
it is necessary to check the two points:

1)the space of admissible local functionals should be closed under
the Poisson bracket;

2)the Jacobi identity should be fulfilled for arbitrary
admissible functionals;

and this check is not explicitly demonstrated in the cited works.
For infinite domain the first requirement was studied by Brown and
Henneaux [BH]. In the finite domain case the important contribution was
made by Lewis, Marsden, Montgomery and Ratiu [LMMR] who showed that the
standard bracket did not fulfil the Jacobi identity and proposed
to modify it by adding some surface terms. Unfortunately these authors
do not consider explicitly the first requirement.
We hope that our results
could serve as a generalization of their ansatz and could be useful
in dealing with interesting problems of field theory.

\section{6. Acknowledgements.}
The author is most grateful to Professor J.Kijowski for the
invitation and to the Stephan Banach Center for hospitality and support.

Discussions with M.Asorey, J.Jezierski, I.Kanatchikov, J.Kijowski,
J.Louko, J.Nester and other participants of Symposium on Differential
Geometry and Mathematical Physics are gratefully acknowledged.

\references{sol93}{

\item{[Ald]}
S.J. \spa{Aldersley},
{\it Higher Eulerian operators and some of their applications}\/,
J.~Math. Phys.  20 (1979), 522-531.

\item{[And76]}
I.M. \spa{Anderson},
{\it Mathematical foundations of the Einstein field equations}\/,
Ph.D.~thesis, Univ. of Arizona, 1976;

\item{[And78]}
I.M. \spa{Anderson},
{\it Tensorial Euler-Lagrange expressions and conservation laws}\/,
Aequationes Mathematicae 17 (1978), 255-291.

\item{[And92]}
I.M. \spa{Anderson}
{\it Introduction to the variational bicomplex.\rm In:
Mathematical aspects of classical field theory ( Eds. M.J.Gotay,
J.E.Marsden and V.Moncrief,
Contemporary Mathematics 132}\/,  AMS, Providence, 1992.

\item{[Arn]}
V.I. \spa{Arnold},
{\it Mathematical methods of classical mechanics}\/,
Nauka, Moscow, 1974 (in Russian).

\item{[ADM]}
R. \spa{Arnowitt}, S. \spa{Deser} and C.W. \spa{Misner},
{\it Consistency of the canonical reduction of General Relativity}\/,
J.~Math. Phys. 1 (1960), 434-439.

\item{[BH]}
J.D. \spa{Brown} and M. \spa{Henneaux},
{\it On the Poisson brackets of differential generators in classical
field theory}\/,
J.~Math. Phys. 27 (1986), 489-491.

\item{[Dorf]}
I. \spa{Dorfman}
{\it Dirac Structures and Integrability of Nonlinear
Evolution Equations}\/, John Wiley and Sons, New York, 1993.

\item{[GD]}
I.M. \spa{Gel'fand} and L.A. \spa{Dickey},
{\it Asymptotics of Sturm-Liouville equation resolvent and algebra of
Korteweg-de Vries equation}\/,
Uspekhi Matematicheskikh Nauk 30 (1975), 67-100 (in Russian).

\item{[JK]}
J. \spa{Jezierski} and J. \spa{Kijowski},
{\it The localization of energy in gauge field theories and in linear
gravitation}\/,
Gen. Rel. Grav. 22 (1990), 1283-1307.

\item{[KT]}
J. \spa{Kijowski} and W.M. \spa{Tulczyjew},
{\it A symplectic framework for field theories}\/,
Lect. Notes in Phys. 107, Springer, New York, 1979.

\item{[KMGZ]}
M.D. \spa{Kruskal}, R.M. \spa{Miura}, C.S. \spa{Gardner} and
N.J. \spa{Zabusky},
{\it Korteweg-de Vries equation and generalizations. V.Uniqueness and
nonexistence of polynomial conservation laws}\/,
J.~Math. Phys. 11 (1970), 952-960.

\item{[LMMR]}
D. \spa{Lewis}, J. \spa{Marsden}, R. \spa{Montgomery} and T. \spa{Ratiu},
{\it The Hamiltonian structure for dynamic free boundary problems}\/,
Physica D18 (1986), 391-404.

\item{[LR]}
A.N. \spa{Leznov}, A.V. \spa{Razumov},
{\it The canonical symmetry for integrable systems}\/,
J. Math. Phys. 35 (1994), 1738-1754.

\item{[Nij]}
A. \spa{Nijenhuis},
{\it Jacobi-type identities for bilinear differential concomitants
of certain tensor fields. I}\/,
Indagationes Math. 17 (1955), 390-397.

\item{[Olv84]}
P.J. \spa{Olver},
{\it Hamiltonian perturbation theory and water waves. \rm In:
Fluids and Plasmas: Geometry and Dynamics, J.E.Marsden (ed.),
Contemporary Mathematics,  28)}\/, AMS, Providence, 1984.

\item{[Olv86]}
P.J. \spa{Olver},
{\it Applications of Lie Groups to Differential Equations}\/,
Graduate texts in mathematics, Springer-Verlag, New York, 1986.

\item{[RT]}
T. \spa{Regge} and C. \spa{Teitelboim},
{\it Role of surface integrals in Hamiltonian formalism of General
Relativity}\/,
Annals of Physics 88 (1974), 286-318.

\item{[Sol85]}
V.O. \spa{Soloviev},
{\it Algebra of asymptotic Poincar\'e group generators in General
Relativity}\/,
Teor. Mat. Fiz. 65 (1985), 400-415.

\item{[Sol92]}
V.O. \spa{Soloviev},
{\it How canonical are Ashtekar's variables?}\/,
Phys. Lett.  B292 (1992), 30-34.

\item{[Sol93]}
V.O. \spa{Soloviev}, {\it Boundary values as Hamiltonian variables.
I.~New Poisson brackets}\/,
J.~Math. Phys.  34 (1993), 5747-5769.

\item{[Sol94]}
V.O. \spa{Soloviev},
{\it Boundary values as Hamiltonian variables. II.~Graded structures}\/,
q-alg/9501017, Preprint IHEP 94-145, Protvino, 1994.

}
\end{document}